\documentclass[longbibliography,twocolumn,superscriptaddress,floatfix,showpacs,nofootinbib]{revtex4-1}
\usepackage{amsfonts, amsmath, amssymb,dsfont, physics}
\usepackage[dvipsnames]{xcolor}
\usepackage{graphicx}
\usepackage{tabularx}
\usepackage{hyperref}
\usepackage{tikz}
\usepackage{xstring}
\usepackage{bbold}
\usepackage{ifthen}

\usetikzlibrary{calc, decorations.markings}
\usetikzlibrary {shapes.geometric}
\newcommand{\mc}[1]{\mathcal{#1}}
\newcommand{\mb}[1]{\mathbb{#1}}
\def\lact  {\,{\triangleright}\,}
\def\ract  {\,{\triangleleft}\,}
\newcommand{\F}[1]{
	\IfEqCase{#1}{
		{0}{{}^{\mc{D}_1}\!F}
		{1}{{}^\triangleright\!F}
		{2}{{}^{\bowtie}\!F}
		{3}{{}^\triangleleft\!F}
		{4}{{}^{\mc{D}_2}\!F}
		{x}{{}^x\!F}
	}
}
\newcommand{\Fi}[1]{
	\IfEqCase{#1}{
		{0}{{}_{\mc{D}_1}\mspace{-1mu}F}
		{1}{{}_\triangleright\mspace{-1mu}F}
		{2}{{}_{\bowtie}\mspace{-1mu}F}
		{3}{{}_\triangleleft\mspace{-1mu}F}
		{4}{{}_{\mathcal{D}_2}\mspace{-1mu}F}
		{x}{{}_x\mspace{-1mu}F}
	}
}

\definecolor{UGentBlue}{RGB}{30,100,200}
\definecolor{UGentYellow}{RGB}{255,210,0}
\definecolor{DarkGray}{RGB}{102, 102, 102}

\colorlet{D1}{black}
\colorlet{D2}{Orange}
\colorlet{M}{UGentBlue}
\colorlet{lattice}{DarkGray}
\colorlet{physical}{purple}
\colorlet{mult}{black!10}

\hypersetup{
	colorlinks,
	linkcolor={UGentBlue},
	citecolor={UGentBlue},
	urlcolor={UGentBlue}
}
\newcommand{\toricTube}[9]{
\begin{tikzpicture}[thick,
	baseline={([yshift=-.5ex]current bounding box.center)},
	scale=.9
	]

\ifthenelse{\equal{#8}{M}}{
    \ifthenelse{\equal{#9}{1}}{
        \begin{scope}[line cap=round,decoration={markings,mark=at position 0.6 with {\arrow{stealth}}}]
        	\draw[#6, postaction={decorate}] (-.25,-.5) -- (.28,-.5);
        	\draw[#6, postaction={decorate}] (-.25,-1) -- (.28,-1);
        	\draw[#7, postaction={decorate}] (.26,-.7) --++ (.53,0);
        	\draw[#7, postaction={decorate}] (.36,-1.2) --++ (.53,0);
        \end{scope}
        \node[#6] at (-.4,-.5) {$#1$};
        \node[#6] at (-.4,-1.) {$#2$};
        \node[#7] at (1,-.7) {$#3$};
        \node[#7] at (1.1,-1.2) {$#4$};
        \node[#8] at (0.1,0) {$#5$};
        \begin{scope}[decoration={markings,mark=at position {#9*0.1} with
        {\arrow{stealth}}}]
        	\draw[#8,dotted, postaction={decorate}] (.25,-.75) arc ({#9*180}:{#9*-180}:{sqrt(3)/2*1.5-.25});
        \end{scope}
        }{
         \begin{scope}[line cap=round,decoration={markings,mark=at position 0.6 with {\arrow{stealth}}}]
    		\draw[D2, postaction={decorate}] (-.29,-.75) --++ (.53,0);
    		\draw[D2, postaction={decorate}] (-.19,-1.2) --++ (.53,0);
    		\draw[D1, postaction={decorate}] (.29,-.5) --++ (.53,0);
    		\draw[D1, postaction={decorate}] (.29,-1) --++ (.53,0);
        \end{scope}
        \node[#6] at (-.44,-.75) {$#1$};
        \node[#6] at (-.4,-1.2) {$#2$};
        \node[#7] at (1,-.5) {$#3$};
        \node[#7] at (1,-1.) {$#4$};
        \node[#8] at (0.1,0) {$#5$};
        \begin{scope}[decoration={markings,mark=at position {#9*0.1} with
        {\arrow{stealth}}}]
        	\draw[#8,dotted, postaction={decorate}] (.25,-.75) arc ({#9*180}:{#9*-180}:{sqrt(3)/2*1.5-.25});
        \end{scope}
        
        }
    }{
    \begin{scope}[line cap=round,decoration={markings,mark=at position 0.6 with {\arrow{stealth}}}]
    	\draw[#6, postaction={decorate}] (-.25,-.5) -- (.28,-.5);
    	\draw[#6, postaction={decorate}] (-.25,-1) -- (.28,-1);
    	\draw[#7, postaction={decorate}] (.26,-.7) --++ (.53,0);
    	\draw[#7, postaction={decorate}] (.36,-1.2) --++ (.53,0);
    \end{scope}
    \begin{scope}[decoration={markings,mark=at position {#9*0.1} with
    {\arrow{stealth}}}]
    	\draw[#8,solid, postaction={decorate}] (.25,-.75) arc ({#9*180}:{#9*-180}:{sqrt(3)/2*1.5-.25});
    \end{scope}
    \node[#6] at (-.4,-.5) {$#1$};
    \node[#6] at (-.4,-1.) {$#2$};
    \node[#7] at (1,-.7) {$#3$};
    \node[#7] at (1.1,-1.2) {$#4$};
    \node[#8] at (0.1,0) {$#5$};
}

\end{tikzpicture}
}

\begin{document}
	
\title{Mapping between Morita equivalent string-net states \texorpdfstring{\\}{}with a constant depth quantum circuit}
\author{Laurens~\surname{Lootens}}\email{laurens.lootens{@}ugent.be}
\affiliation{\it Department of Physics and Astronomy, Ghent University, Krijgslaan 281, 9000 Gent, Belgium}
\author{Bram~\surname{Vancraeynest-De Cuiper}}
\affiliation{\it Department of Physics and Astronomy, Ghent University, Krijgslaan 281, 9000 Gent, Belgium}
\author{Norbert~\surname{Schuch}}
\affiliation{University of Vienna, Faculty of Mathematics, Oskar-Morgenstern-Platz 1, 1090 Wien, Austria,
and\\
University of Vienna, Faculty of Physics, Boltzmanngasse 5, 1090 Wien, Austria}
\author{Frank~\surname{Verstraete}}
\affiliation{\it Department of Physics and Astronomy, Ghent University, Krijgslaan 281, 9000 Gent, Belgium}

\begin{abstract}
We construct a constant depth quantum circuit that maps between Morita equivalent string-net models.  Due to its constant depth and unitarity, the circuit cannot alter the topological order, which demonstrates that Morita equivalent string-nets are in the same phase. The circuit is constructed from an invertible bimodule category connecting the two input fusion categories of the relevant string-net models, acting as a generalized Fourier transform for fusion categories. The circuit does not only act on the ground state subspace, but acts unitarily on the full Hilbert space when supplemented with ancillas.
\end{abstract}

\maketitle

%%%%%%%%%%%%%%%%%%%%%%%%%%%%%%%%%%%%%%%%%%%%%%%%%%%%%%%%%%%%%%%%%%%%%%%%%%%%%%%%%%%%%%%%

\section{Introduction}

Ever since their conception, the string-net models as originally proposed by Levin and Wen \cite{levin2005string} and their subsequent generalizations \cite{kitaev2012models,lan2014topological,lin2014generalizations,lake2016signatures,hahn2020generalized,lin2021generalized} have provided a rich playground for studying microscopic realisations of non-chiral topologically ordered phases of matter in 2+1 dimensions. Taking a unitary fusion category (UFC) $\mc{D}$ \cite{etingof2016tensor} as input, these exactly solvable models allow for the explicit realization of several key features of topologically ordered systems, such as ground state degeneracies that depend on the topology of the space and anyonic quasi-particle excitations that satisfy non-trivial braiding statistics. From the category-theoretical side, the different ground states and excitations are described by the monoidal center $Z(\mc{D})$ of the input UFC $\mc{D}$ \cite{kitaev2012models}, which is itself a unitary modular fusion category (UMFC) and describes the topological order. The fact that the UMFC describing the topological order is always the center of some other UFC is the reason why string-net models are only able to describe non-chiral topological order with gappable boundaries; exactly solvable lattice models for chiral topological order have proven more challenging to obtain \cite{kapustin2020local}.\\

An important observation is that the process of going to the center $Z(\mc{D})$ of a UFC $\mc{D}$ is not an injective operation, or put differently that distinct UFCs $\mc{D}_1$ and $\mc{D}_2$ can have the same center $Z(\mc{D}_1) \simeq Z(\mc{D}_2)$ \cite{kitaev2012models}. This has lead to the conjecture that two such string-net models based on different $\mc{D}_1$ and $\mc{D}_2$ belong to the same topological phase, despite the fact that microscopically they can look very different. UFCs with the same center are said to be Morita equivalent \cite{muger2003subfactors}, and the collection of UFCs Morita equivalent to $\mc{D}$ is called its Morita class. A direct consequence of Morita equivalence is that the topological excitations or anyons of two Morita equivalent string-net models are in one-to-one correspondence to each other. Kitaev and Kong demonstrated the existence of an invertible domain wall between two Morita equivalent string-net models through which anyons can move freely without condensing on the domain wall \cite{kitaev2012models}. Mathematically, the existence of such an invertible domain wall is guaranteed by the fact that for any two Morita equivalent UFCs $\mc{D}_1$ and $\mc{D}_2$ there exists an invertible $(\mc{D}_1,\mc{D}_2)$-bimodule category $\mc{M}$ \cite{etingof2016tensor}. The central goal of this paper is to show that this implies the existence of a constant depth quantum circuit which maps the different string-net states into each other.\\

From a quantum information point of view, it has been understood that two states are in the same phase if there exists a constant depth quantum circuit that is able to map between the two \cite{hastings2005quasiadiabatic,PhysRevLett.97.050401,osborne2007simulating}. In this work, we provide such a constant depth circuit for the case of Morita equivalent string-nets $\mc{D}_1$ and $\mc{D}_2$ that maps the ground states as well as the excitations of one model to those of the other. For the ground state, the quantum circuit can be understood as a generalized version of the Hamiltonian, using the invertible $(\mc{D}_1,\mc{D}_2)$-bimodule category $\mc{M}$ to intertwine the Hamiltonian of the string-net $\mc{D}_1$ to the Hamiltonian of the string-net $\mc{D}_2$. Our construction generalizes a previous mapping, that was obtained from Kitaev's quantum double models \cite{kitaev2003fault} for a group $G$ to string-net models $\text{Rep}(G)$ \cite{buerschaper2009mapping,kadar2010microscopic}, to the general case of two Morita equivalent string-nets.\\

\emph{Outline:} we begin by providing a brief review of string-nets, focusing mainly on a generalization known as the extended string-net models \cite{hu2018full}. The reason for this is that these generalizations allow for a more unified treatment of the vertex and plaquette excitations, the two types of excitations in a string-net model. They are understood in the diagrammatic language as idempotents of the tube algebra associated to the string-net UFC $\mc{D}$ \cite{lan2014topological}, which is well known to provide a characterization of the center $Z(\mc{D})$. Next, we discuss how an invertible $(\mc{D}_1,\mc{D}_2)$-bimodule category can be used to provide a map between the tube algebras of the UFCs $\mc{D}_1$ and $\mc{D}_2$, and thereby show how excitations from one model are mapped to excitations of the other. Taking all this together, we are able to explicitly construct a constant depth unitary quantum circuit which we explain in detail.\\

For the case of a hexagonal lattice,  this circuit is depth three, which follows from the fact that our circuit sequentially acts on three sublattices. This is depicted in Figure ~\ref{fig:QuantumCircuit3D} below. The action of the circuit on the three sublattices is worked out in the main text. Each of these local unitaries acts on 18 degrees of freedom associated to a plaquette. We show that the unitary map on the three sublattices is respectively constructed from the different associators of the $\left(\mc{D}_1,\mc{D}_2\right)$-bimodule category $\mc{M}$, which are the solutions to a set of coupled pentagon equations~\cite{etingof2016tensor,lootens2021matrix}. In this way our construction gives a very concrete quantum information perspective on the categorical notion of Morita equivalence. We also show how the quantum circuit can be generalized to act on the full (extended) string-net Hilbert space and obtain a very similar circuit as in the absence of excitations, the details of which are relegated to Appendix \ref{App:ExtendedCircuit}.

\begin{figure}[htb]
	\centering
	\input{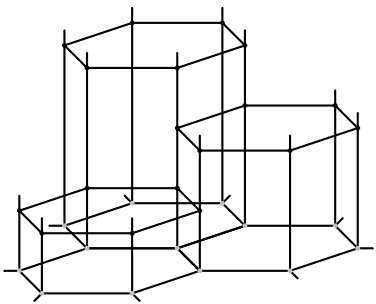}
	\caption{An illustration of how our circuit acts on the three sublattices of the hexagonal lattice.}
	\label{fig:QuantumCircuit3D}
\end{figure}

\section{String-nets}
In this section we give a brief review of Levin and Wen's string-net models on the hexagonal lattice. First, we review the ordinary string-net model, hereby closely following the original work of Levin and Wen~\cite{levin2005string}, and then we turn to the extended string-nets \cite{hu2018full}. These can be regarded as a generalization of the ordinary string-nets that provide a more natural and convenient setting in which to study and classify excited states.

\subsection{Levin and Wen's string-nets}

A string-net model is defined by the data of some given input unitary fusion category $\mc{D}$. This input data contains a set of simple objects (string types) that we will denote by $a,b,c,...$ with corresponding fusion rules $a\otimes b \simeq \bigoplus_c N^c_{ab} c$, corresponding quantum dimensions $d_a,d_b,d_c,...$ and an isomorphism $F: (a \otimes b) \otimes c \rightarrow a \otimes (b \otimes c)$. Given some oriented trivalent lattice, such as the hexagonal lattice in this work, the Hilbert space of the string-net is defined by configurations of the string types living on the edges of the lattice. Furthermore, for every string type $a$ there exists a corresponding conjugate string type $\overline{a}$ such that $\overline{\overline{a}} \equiv a$ and $\overline{a}$ can be interpreted as $a$ living on the same edge with the opposite orientation. The quantum dimensions obey $d_{\overline{a}}=d_a$. There is a unique self-conjugate vacuum string \textbf{1}, $\overline{\textbf{1}}\equiv\textbf{1}$.\\

The string-net Hamiltonian consists of two terms:
\begin{equation}
	H = -\sum_v A_v - \sum_p B_p,
\end{equation}
Herein the two sums in the Hamiltonian are over the vertices and plaquettes of the lattice respectively. The sum over $s$ is a sum over all string types of the input UFC. $D$ is the total quantum dimension defined as $D = \sqrt{\sum_i d_i^2}$. The vertex operators $A_v$ act on the vertices according to
\begin{equation}
	A_v\ket{\input{vertex_convention.tikz}} = \delta_{a,\overline{b},\overline{c}}\ket{\input{vertex_convention.tikz}}
\end{equation}
where $\delta_{a,b,c}$ is $1$ whenever the fusion of $a\otimes b$ contains $\overline{c}$ and is $0$ else. The plaquette operators $B_p$ are defined as
\begin{equation}
	B_p = \frac{1}{D}\sum_a d_a B_p^a,
\end{equation}
where the terms $B_a^s$ acts on a plaquette by inserting a clockwise oriented string of type $a$ in the plaquette and fusing it to the lattice. This is done by making repeated use of the resolution of the identity for UFCs, namely,
\begin{equation}
	\input{res_id_1.tikz}
	=
	\sum_{c,n}\sqrt{\frac{d_c}{d_ad_b}}
	\input{res_id_2.tikz},
	\label{eq:res_id}
\end{equation}
the $F$-moves of the given input UFC
\begin{equation}
	\input{0F_2.tikz}
	= \sum_{f,mn}\left(F^{abc}_d\right)^{f,mn}_{e,jk}
	\input{0F_1.tikz},
\end{equation}
and the ``bubble pop'' identity
\begin{equation}
	\input{del_bubble_1.tikz}
	=
	\sqrt{\frac{d_ad_b}{d_c}}\delta_{m,n}\delta_{c,c'}
	\begin{tikzpicture}[thick,decoration={
			markings,
			mark=at position 0.6 with {\arrow{stealth}}},
		baseline={([yshift=-.5ex]current bounding box.center)},
		line cap=round
		] 
		
\begin{scope}[black]
	\draw[D1, postaction={decorate}] (0,-.6) -- (0,1.6);
\end{scope}
\node[D1] at (0,-.8) {$c$};
\node[D1] at (0,1.8) {$c$};
\end{tikzpicture}.
	\label{eq:del_bubble}
\end{equation}
The matrix elements of this operator are worked out in detail in~\cite{levin2005string,hahn2020generalized}. Every operator $B_p^a$ is a 18-body operator in the sense that its action explicitly depends on the six string types living on the edges of the plaquette $p$, the six multiplicities and the six edges adjacent to the plaquette, but acts diagonally on the latter. It can be shown that both $A_v$ and $B_p$ are Hermitian projection operators, $A_v^2=A_v$ and $B_p^2=B_p$, that moreover all commute. Hence, the string-net condensed RG fixed point ground states are the simultaneous eigenvectors of all $A_v$ and $B_p$. Quasiparticle excitations on top of these ground states are obtained by violating at least one of the projector constraints. These excitations are gapped and necessarily come in pairs. Vertex - or electric excitations are obtained by violating the $A_v$ constraints whereas plaquette - or magnetic excitations violate the $B_p$ constraints. These excitations are treated in a unified way in the language of extended string-nets which we now revisit.

\subsection{Extended string-nets}

The extended string-net models \cite{hu2018full} are very similar to the ordinary string-net models except for a few key differences which we discuss in this section.\\

\begin{figure}
	\centering
	\input{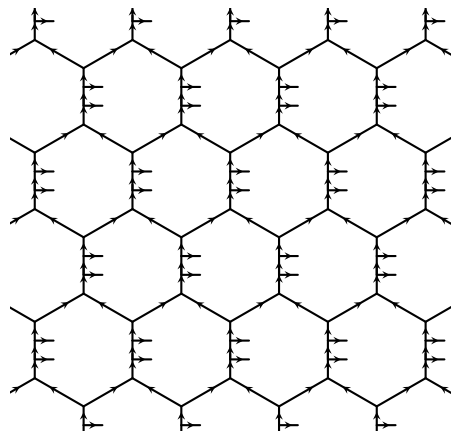}
	\caption{We adapt the convention that all arrows on the edges point upward and the open legs are attached to the lattice as depicted where every open edge corresponds to the closest vertex of the hexagonal lattice. Multiplicity degrees of freedom are suppressed.}
	\label{fig:ExtendedHex}
\end{figure}

Given a hexagonal lattice we associate with every vertex an additional open edge that is attached to an edge that is connected to the vertex under consideration. These open edges also carry string degrees of freedom. The choice of which open edge belongs to which vertex is to some extent arbitrary and the open edges can even be moved around the plaquette by a series of $F$-moves but we will adapt the convention denoted in Figure~\ref{fig:ExtendedHex}. Furthermore we demand the fusion rules to be satisfied on every vertex, $\delta_{a,b,c} = 1$. Finally, the Hamiltonian of the extended string-net is of the same form as the Hamiltonian of the ordinary string-net model:
\begin{equation}
	H = -\sum_v \mc{A}_v -\sum_p \mc{B}_p,
\end{equation}
where the $\mc{A}_v$ and the $\mc{B}_p$ are again mutually commuting Hermitian projectors. $\mc{A}_v$ acts on vertices according to
\begin{equation}
	\mc{A}_v\ket{\input{vertex_convention_extended.tikz}} = \delta_{d,\textbf{1}}\ket{\input{vertex_convention_extended.tikz}}.
\end{equation}
The action of $\mc{B}_p$ on the plaquette is explained in detail in~\cite{hu2018full} and reduces for $p_1, p_2=\textbf{1}$ to the action of the plaquette operator $B_p$ of the ordinary model, where $p_1, p_2$ denote the open edges of the plaquette $p$.\\

The extended string-net model thus deals with vertex excitations in the ordinary string-net model by considering extra degrees of freedom that signal the presence of vertex projector violations.

\subsection{Tube algebra}

In the extended string-net models, vertex and plaquette excitations are both characterized by the \textit{tube algebra} \cite{lan2014topological,hu2018full}. The action of a tube $\mathcal{T}_i$ on a plaquette can graphically be depicted as the insertion of the following diagram in a plaquette:
\begin{equation}
	\input{tube_black.tikz}
\end{equation}
and fusing it to the boundary; here, $i$ runs over all possible configurations that give a nonzero tube. The set of tubes is closed under multiplication, 

\begin{equation}
	\mc{T}_i\mc{T}_j
	= \sum_k f_{ij}^k \mc{T}_k,
\end{equation}
defined by stacking:

\begin{equation}
	\input{tube_black_mult_ij.tikz}
	= \sum_k f_{ij}^k \input{tube_black_mult_k.tikz},
\end{equation}

as well as Hermitian conjugation:
\begin{equation}
	\left(\mc{T}_i\right)^\dag = \sum_j u_i^j \mc{T}_j
\end{equation}
\cite{bultinck2017anyons,williamson2017symmetry}. This turns the tube algebra into a finite $C^\star$~-~algebra, for which the Artin-Wedderburn theorem dictates that one can find an isomorphism to a direct sum of simple matrix algebras. Explicitly, this implies the existence of simple idempotents $p_{ii}^a$ and nilpotents $p_{ij}^a, i \neq j$ that satisfy

\begin{equation}
	\begin{gathered}
		p_{ij}^a = \sum_k x_{ij}^{a,k} \mc{T}_k, \quad p_{ij}^a p_{kl}^b = \delta_{ab} \delta_{jk} p_{il}^a,\\ \left(p_{ij}^a\right)^\dag = p_{ji}^a.
	\end{gathered}
\end{equation}

The elementary excitations of the extended string-net model correspond to the simple idempotents $p_{ii}^a$ of the tube algebra, each of which corresponds to a specific combination of one plaquette excitation and two vertex excitations. These simple idempotents have to be grouped into central idempotents $P_a$, satisfying
\begin{equation}
	P_a = \sum_{i=0}^{n_a-1} p_{ii}^a, \quad [P_a,\mathcal{T}_i] = 0, \quad \sum_a P_a = \mathds{1},
\end{equation}
where $n_a$ denotes the dimension of the central idempotent $P_a$. These central idempotents then correspond to the irreducible representations of the tube algebra, and are in one-to-one correspondence with the simple objects of the monoidal center $Z(\mc{D})$.

\section{Bimodule categories}
We will consider two string-net models $\mc{D}_1$ and $\mc{D}_2$ that are supposedly in the same phase; this requires that $\mc{D}_1$ and $\mc{D}_2$ are Morita equivalent, $Z(\mc{D}_1) \simeq Z(\mc{D}_2)$. From a categorical perspective, this implies the existence of an invertible $(\mc{D}_1,\mc{D}_2)$-bimodule category $\mc{M}$. In this section, we briefly review some properties of a bimodule category, mainly to fix notation \footnote{The $\F{0}$, $\F{1}$, $\F{2}$, $\F{3}$ and $\F{4}$ symbols in this work are denoted as ${}^0\!F$, ${}^1\!F$, ${}^2\!F$, ${}^3\!F$ and ${}^4\!F$ symbols in~\cite{lootens2021matrix} respectively.}, and show how this bimodule category provides a map between the excitations of the two string-net models.

\subsection{Bimodule categories}
We start with a spherical UFC $\mc{D}_1$, with objects $a,b,c,... \in \mc{D}_1$ and an associator expressed in a basis of simple objects as
\begin{equation}
	\input{0F_1.tikz}
	= \sum_{f,mn}\left(\F{0}^{abc}_d\right)^{f,mn}_{e,jk}
	\hspace{-16pt}
	\input{0F_2.tikz},
	\label{eq:0F}
\end{equation}
where in the diagrams we will color lines labeled by objects in $\mc{D}_1$ in black for the remainder of this work. We also have a secondv spherical UFC $\mc{D}_2$, with objects $\alpha,\beta,\gamma,... \in \mc{D}_2$ and an associator given by
\begin{equation}
	\input{4F_1.tikz}
	\hspace{-12pt}
	= \sum_{\nu,mn}\left(\F{4}^{\alpha\beta\gamma}_\delta\right)^{\nu,mn}_{\mu,jk}
	\hspace{-16pt}
	\input{4F_2.tikz},
	\label{eq:4F}
\end{equation}
with lines labeled by objects in $\mc{D}_2$ drawn in orange. A left $\mc{D}_1$-module category $\mc{M}$ is a category $\mc{M}$ with objects $A,B,C,... \in \mc{M}$, a left-action of $\mc{D}_1$ on $\mc{M}$ given by $a \lact A \simeq \bigoplus_B N_{aA}^B B$ and an isomorphism $\F{1}: (a \otimes b) \lact A \rightarrow a \lact (b \lact A)$, which can graphically be expressed as
\begin{equation}
	\input{1F_1.tikz}
	= \sum_{C,mn}\left(\F{1}^{abA}_B\right)^{C,mn}_{c,jk}
	\input{1F_2.tikz},
	\label{eq:1F}
\end{equation}
where lines labeled by objects in $\mc{M}$ are blue. Similarly, a right $\mc{D}_2$-module category $\mc{M}$ has a right-action of $\mc{D}_2$ on $\mc{M}$ given by $A \ract \alpha \simeq \bigoplus_B N_{A\alpha}^BB$ and an isomorphism $\F{3}: (A \ract \alpha) \ract \beta \rightarrow A \ract (\alpha \otimes \beta)$, graphically expressed as
\begin{equation}
	\input{3F_1.tikz}
	= \sum_{\gamma,mn}\left(\F{3}^{A\alpha\beta}_B\right)^{\gamma,mn}_{C,jk}
	\input{3F_2.tikz}.
	\label{eq:3F}
\end{equation}
A $(\mc{D}_1,\mc{D}_2)$-bimodule category $\mc{M}$ is then defined as a left-$\mc{D}_1$ and right-$\mc{D}_2$ module category $\mc{M}$ that additionally is equipped with an isomorphism $\F{2}: (a \lact A) \ract \alpha \rightarrow a \lact (A \ract \alpha)$, graphically depicted as
\begin{equation}
	\input{2F_1.tikz}
	\hspace{-12pt}
	= \sum_{D,mn}\left(\F{2}^{a A \alpha}_B\right)^{D,mn}_{C,jk}
	\hspace{-18pt}
	\input{2F_2.tikz}.
	\label{eq:2F}
\end{equation}
Physically, these bimodule categories arise most evidently in the study of domain walls between different string-net models \cite{kitaev2012models}, where the various associativity conditions discussed above amount to the generalizations of the bulk $F$-moves. For a general domain between Morita equivalent string-net models, the process of pushing excitations through such a domain wall is not reversible, as several excitations can condense on the domain wall. In the special case where excitations can freely move through the domain wall, we are dealing with an \textit{invertible} bimodule category. For every two Morita equivalent UFCs $\mc{D}_1$ and $\mc{D}_2$ one can always write down an invertible $(\mc{D}_1,\mc{D}_2)$-bimodule category, and we will use this particular bimodule category in the remainder of this work.\\

The $(\mc{D}_1,\mc{D}_2)$-bimodule category $\mc{M}$ does not have an intrinsic duality, but one can define a $(\mc{D}_2,\mc{D}_1)$-bimodule category denoted by $\mc{M}^\text{op}$ or $\overline{\mc{M}}$ to contain the dual objects of $\mc{M}$. Crucially, this allows us to impose a generalized notion of sphericality \cite{kitaev2012models,schaumann2013traces}, which we will need to perform certain diagrammatical manipulations. Concretely, invertibility of $\mc{M}$ can be used to generalize the resolution of the identity (\ref{eq:res_id}) of UFCs to a resolution of the identity which allows for the creation of a $\mc{D}_1$- or $\mc{D}_2$-line from the fusion of an $\mc{M}$-line with an $\mc{M}^\text{op}$-line:
\begin{equation}
	\input{M_x_Mop_1.tikz}
	=
	\sum_{a,n}\sqrt{\frac{d_a}{d_Ad_B}}
	\input{M_x_Mop_2.tikz},
	\label{bimod_inv1}
\end{equation}
\begin{equation}
	\input{Mop_x_M_1.tikz}
	=
	\sum_{\alpha, n}\sqrt{\frac{d_\alpha}{d_Ad_B}}
	\input{Mop_x_M_2.tikz}.
	\label{bimod_inv2}
\end{equation}
\subsection{Tube algebra bimodules}

Using the bimodule categories $\mc{M}$ and $\overline{\mc{M}}$ we can write down bimodule tubes for the tube algebras $\mc{T}^{\mc{D}_1}$ and $\mc{T}^{\mc{D}_2}$ of $\mc{D}_1$ and $\mc{D}_2$ respectively. These bimodule tube algebras, denoted by $\mc{T}^\mc{M}$ and $\mc{T}^{\overline{\mc{M}}}$, are generated by
\begin{equation}
	\mc{T}_i^\mc{M} = \input{tube.tikz}, \quad
	\mc{T}_i^{\overline{\mc{M}}} = \input{tube_op.tikz}.
\end{equation}
They satisfy
\begin{align}
	\mc{T}_i^\mc{M} \mc{T}_j^{\mc{D}_1} = \sum_k a_{ij}^k \mc{T}_k^\mc{M}, \quad
	\mc{T}_i^{\mc{D}_2} \mc{T}_j^\mc{M} = \sum_k b_{ij}^k \mc{T}_k^\mc{M},\\
	\mc{T}_i^{\mc{D}_1} \mc{T}_j^{\overline{\mc{M}}} = \sum_k c_{ij}^k \mc{T}_k^{\overline{\mc{M}}}, \quad
	\mc{T}_i^{\overline{\mc{M}}} \mc{T}_j^{\mc{D}_1} = \sum_k d_{ij}^k \mc{T}_k^{\overline{\mc{M}}},\\
	\mc{T}_i^{\overline{\mc{M}}} \mc{T}_j^\mc{M} = \sum_k e_{ij}^k \mc{T}_k^{\mc{D}_1}, \quad
	\mc{T}_i^\mc{M} \mc{T}_j^{\overline{\mc{M}}} = \sum_k g_{ij}^k \mc{T}_k^{\mc{D}_2},
	\label{tube_M_Mop}
\end{align}
where the last line requires the invertibility of the bimodule category, manifested in the identities \eqref{bimod_inv1} and \eqref{bimod_inv2}. These bimodule tubes are mapped into one another under Hermitian conjugation:
\begin{equation}
	\left(\input{dagger_tube_1.tikz}\right)^\dagger
	= \sqrt{\frac{d_{c_1}d_{c_2}}{d_{\gamma_1}d_{\gamma_2}}}
	\input{dagger_tube_2.tikz}.
	\label{eq:ConjugateTube}
\end{equation}
This Hermitian conjugation is defined such that the operators obtained from inserting $\mc{T}_i^{\mc{M}}$ and $(\mc{T}_i^{\mc{M}})^\dag$ in a plaquette expressed as matrices acting on the relevant degrees of freedom are related via the usual Hermitian conjugation. Using these properties, we can define a big $C^\star$~-~algebra generated by $\{\mc{T}_i^{\mc{D}_1},\mc{T}_j^\mc{M},\mc{T}_k^{\overline{\mc{M}}},\mc{T}_l^{\mc{D}_2}\}$, where we take undefined tube multiplications to be zero. This big $C^\star$~-~algebra contains the $C^\star$~-~algebras $\mc{T}^{\mc{D}_1}$ and $\mc{T}^{\mc{D}_2}$ as subalgebras. Using the Artin-Wedderburn theorem for this big (finite) $C^\star$~-~algebra, there exists an isomorphism to a direct sum of simple matrix algebras, the number of which equals the number of central idempotents of $\mc{T}^{\mc{D}_1}$ and $\mc{T}^{\mc{D}_2}$. In addition to the simple idempotents and nilpotents of $\mc{T}^{\mc{D}_1}$ and $\mc{T}^{\mc{D}_2}$, we can define simple bimodules:
\begin{gather}
	p_{ij}^{a,\mc{M}} = \sum_{k} y_{ij}^{a,k} \mc{T}_k^\mc{M}, \quad p_{ji}^{a,\overline{\mc{M}}} = \sum_{k} z_{ji}^{a,k} \mc{T}_k^{\overline{\mc{M}}},\\
	\left(p_{ij}^{a,\mc{M}}\right)^\dag = p_{ji}^{a,\overline{\mc{M}}},
\end{gather}
with $0 \leq i < n_a^{\mc{D}_2}$ and $0 \leq j < n_a^{\mc{D}_1}$, such that they satisfy
\begin{align}
    \nonumber
	\hspace{-3pt} p_{ij}^{a,\mc{D}_1} p_{kl}^{b,\mc{D}_1} &= \delta_{ab} \delta_{jk} p_{il}^{a,\mc{D}_1}, \ 
	p_{ij}^{a,\mc{D}_2} p_{kl}^{b,\mc{D}_2} = \delta_{ab} \delta_{jk} p_{il}^{a,\mc{D}_2},\\
    \nonumber
	\hspace{-3pt} p_{ij}^{a,\mc{M}} p_{kl}^{b,\mc{D}_1} &= \delta_{ab} \delta_{jk} p_{il}^{a,\mc{M}}, \ 
	p_{ij}^{a,\mc{D}_2} p_{kl}^{b,\mc{M}} = \delta_{ab} \delta_{jk} p_{il}^{a,\mc{M}},\\
	\nonumber
	\hspace{-3pt}p_{ij}^{a,\mc{D}_1} p_{kl}^{b,\overline{\mc{M}}} &= \delta_{ab} \delta_{jk} p_{il}^{a,\overline{\mc{M}}}, \ 
	p_{ij}^{a,\overline{\mc{M}}} p_{kl}^{b,\mc{D}_2} = \delta_{ab} \delta_{jk} p_{il}^{a,\overline{\mc{M}}},\\
	\hspace{-3pt}p_{ij}^{a,\overline{\mc{M}}} p_{kl}^{b,\mc{M}} &= \delta_{ab} \delta_{jk} p_{il}^{a,\mc{D}_1}, \ 
	p_{ij}^{a,\mc{M}} p_{kl}^{b,\overline{\mc{M}}} = \delta_{ab} \delta_{jk} p_{il}^{a,\mc{D}_2},
\end{align}
where from now on the range of the indices $i,j,k,l$ should be inferred from the context. The simple bimodules will be the essential building blocks of the quantum circuit we aim to construct, providing a map between the ground states and excitations of the string-net model $\mc{D}_1$ to the ground states and excitations of the string-net model $\mc{D}_2$.

\section{Quantum circuit}

Before we construct our quantum circuit, we have to address the point that the total dimension of the Hilbert space need not match between the string-net model $\mc{D}_1$ and $\mc{D}_2$; the fact that they are Morita equivalent only ensures that number of ground states on some closed manifold is the same. The number of distinct excitations which form a complete basis for the Hilbert space will in general differ between the two models, which is reflected in their tube algebras; since we have $Z(\mc{D}_1) \simeq Z(\mc{D}_2)$, the number of central idempotents $P_a^{\mc{D}_1}$ and $P_a^{\mc{D}_2}$ is the same, but their dimensions $n_a^{\mc{D}_1}$ and $n_a^{\mc{D}_2}$ can differ. This discrepancy in the dimensions of the Hilbert spaces poses an obstacle if we want to construct a unitary quantum circuit that maps the full Hilbert space of one model to the other.\\

To solve this problem, we will place an ancilliary qudit $\ket{i}$ of dimension $n^{\mc{D}_2} = \max\limits_a n_a^{\mc{D}_2}$ in each plaquette of the string-net model $\mc{D}_1$. The original Hilbert space of the string-net model is obtained by fixing these ancillas to $\ket{0}$; we will denote states in this Hilbert space as $\ket{\psi^{\mc{D}_1},0}$. We can then define the following operator:
\begin{equation}
	V_{\mc{D}_1}^{\mc{D}_2} := \left(\sum_{a,ij} p_{ij}^{a,\mc{M}} \otimes\ketbra{j}{i}\right)^{\otimes N}
\end{equation}
with $N$ the number of plaquettes in the system, where $p_{ij}^{a,\mc{M}}$ acts on a plaquette as defined before and $\ketbra{j}{i}$ acts on the corresponding ancilla. The Hermitian conjugate of this operator is given by
\begin{equation}
	\left(V_{\mc{D}_1}^{\mc{D}_2}\right)^\dag = \left(\sum_{a,ij} p_{ij}^{a,\overline{\mc{M}}} \otimes \ketbra{j}{i}\right)^{\otimes N},
\end{equation}
and the product of $V_{\mc{D}_1}^{\mc{D}_2}$ with its Hermitian conjugate is
\begin{align}
	\left(V_{\mc{D}_1}^{\mc{D}_2}\right)^\dag V_{\mc{D}_1}^{\mc{D}_2} &= \left(\sum_{ab,ii'jj'} p_{j'i'}^{a,\overline{\mc{M}}} p_{ij}^{a,\mc{M}} \otimes \ket{i'}\braket{j'}{j}\bra{i} \right)^{\otimes N}\nonumber\\
	&= \left( \sum_{a,ij} p_{jj}^{a,\mc{D}_1} \otimes \ketbra{i} \right)^{\otimes N}\nonumber\\
	&= \left( \sum_{a} P_a^{\mc{D}_1} \otimes \mathds{1}_{n_a^{\mc{D}_2}} \right)^{\otimes N}.
\end{align}
In general this is a projector implying $V_{\mc{D}_1}^{\mc{D}_2}$ is an isometry, but restricting to the original Hilbert space where all ancillas are in the $\ket{0}$ state we get
\begin{align}
	\left(V_{\mc{D}_1}^{\mc{D}_2}\right)^\dag V_{\mc{D}_1}^{\mc{D}_2} \ket{\psi^{\mc{D}_1},0} &= \left( \sum_{a} P_a^{\mc{D}_1}\otimes \mathds{1}_{n_a^{\mc{D}_2}}\right)^{\otimes N} \ket{\psi^{\mc{D}_1},0}\nonumber\\
	&= \ket{\psi^{\mc{D}_1},0},
\end{align}
meaning that the isometry $V_{\mc{D}_1}^{\mc{D}_2}$ is in fact a unitary on the original Hilbert space of the string-net model.\\

\begin{figure}
	\centering
	\input{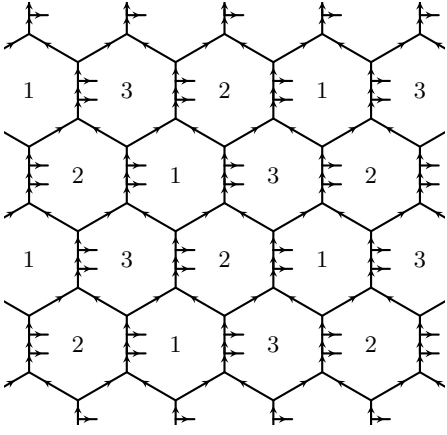}
	\caption{Division of the hexagonal lattice into three sublattices. We build the quantum circuit by acting on the plaquettes in the order denoted in the figure.}
	\label{fig:sublattice}
\end{figure}

A constant depth quantum circuit is obtained by applying the isometry $V_{\mc{D}_1}^{\mc{D}_2}$ in such a way that it is a constant depth operator. For the hexagonal lattice, this can be achieved by dividing the lattice into three sublattices as shown in Figure \ref{fig:sublattice} and acting sequentially on the first, second and third sublattice. This implementation of the isometry $V_{\mc{D}_1}^{\mc{D}_2}$ turns it into a depth three quantum circuit. We now discuss this quantum circuit in more detail, first for the ground states, and then for a generic excited state in the original string-net Hilbert space spanned by $\ket{\psi^{\mc{D}_1},0}$.

\subsection{Ground state}

The ground states of the string-net model $\mc{D}_1$, which we will denote as $\ket{\psi_{\text{GS}}^{\mc{D}_1},0}$, are defined as the eigenvectors of the ground-state projector $P_\text{GS}^{\mc{D}_1}$:
\begin{equation}
	\begin{split}
		P_\text{GS}^{\mc{D}_1} \ket{\psi_{\text{GS}}^{\mc{D}_1},0} &= \left(p_{00}^{0,\mc{D}_1} \otimes \ketbra{0} \right)^{\otimes N}\ket{\psi_{\text{GS}}^{\mc{D}_1},0}\\
		&= \ket{\psi_{\text{GS}}^{\mc{D}_1},0},
	\end{split}
\end{equation}
projecting each plaquette onto the simple idempotent $p_{00}^{0,\mc{D}_1}$ which corresponds to the vacuum; it is equal to the plaquette term of the Hamiltonian,
\begin{equation}
	p_{00}^{0,\mc{D}_1} = \frac{1}{D}\sum_a d_a \mc{B}_p^{a,\mc{D}_1},
\end{equation}
which also projects onto the ground state of the vertex term by fixing the extended string-net legs to the trivial string. At first glance, one would expect the simple idempotent $p_{00}^{0,\mc{D}_1}$ corresponding to the vaccuum to be central; the reason this is not the case is that in the extended string-net model, the simple idempotents keep track of one plaquette and two vertex excitations. If the two vertices are excited with a particle-antiparticle pair then together they are in the vacuum, and the corresponding simple idempotent $p_{ii}^{0,\mc{D}_1}, i \neq 0$ will also be contained in the central idempotent $P_{\mathbb 1}^{\mc{D}_1}$ corresponding to the vacuum in $Z(\mc{D}_1)$.\\

Acting with our quantum circuit on the ground state, we obtain
\begin{equation}
	V_{\mc{D}_1}^{\mc{D}_2} \ket{\psi_{\text{GS}}^{\mc{D}_1},0} = \left( p_{00}^{0,\mc{M}} \otimes \ketbra{0} \right)^{\otimes N} \ket{\psi_{\text{GS}}^{\mc{D}_1},0}.
\end{equation}
One can readily verify that
\begin{align}
	P_\text{GS}^{\mc{D}_2} V_{\mc{D}_1}^{\mc{D}_2} \ket{\psi_{\text{GS}}^{\mc{D}_1},0} &= \left( p_{00}^{0,\mc{M}} \otimes \ketbra{0} \right)^{\otimes N} \ket{\psi_{\text{GS}}^{\mc{D}_1},0},\nonumber\\
	&=\ket{\psi_{\text{GS}}^{\mc{D}_2},0}.
\end{align}
showing that $V_{\mc{D}_1}^{\mc{D}_2}$ does indeed provide a unitary map between the ground states of the string-net model $\mc{D}_1$ and $\mc{D}_2$. The operator $p_{00}^{0,\mc{M}}$ can be interpreted as a generalized plaquette term, in the sense that
\begin{equation}
	p_{00}^{0,\mc{M}} = \frac{1}{D} \sum_A d_A \mc{B}_p^{A,\mc{M}},
\end{equation}
i.e. it is the result of fusing a weighted sum of all loops labeled by $A \in \mc{M}$ with the plaquette. When applied to the string-net configuration where all edges are fixed to the trivial label (a product state), $\left(p_{00}^{0,\mc{M}}\right)^{\otimes N}$ projects onto the ground state of the string-net model $\mc{D}_2$. This operator can be used to derive a projected entangled pair state (PEPS) representation of the string-net ground state, generalizing the constructions in \cite{buerschaper2009explicit,gu2009tensor} to the tensor network representations obtained in \cite{lootens2021matrix}.\\

More explicitly, the application of $\mc{B}_p^{A,\mc{M}}$, $\mc{B}_p^{D,\mc{M}}$ and $\mc{B}_p^{G,\mc{M}}$ to the three sublattices in the order depicted in Figure~\ref{fig:sublattice} can be worked out by a series of recouplings involving the bimodule $F$-symbols. For a plaquette of the first kind, we get
\begin{equation*}
	\input{sublattice_1_a.tikz}
	\hspace{-1.2em}=\hspace{-0.7em}\sum_{\{B_i,k_i\}} \hspace{-0.7em} M^1\hspace{-1.2em}
	\input{sublattice_1_b.tikz}\hspace{-0.7em},
\end{equation*}
where the matrix $M^1$ depends on $\{a_i,A\}$, which we suppress for convenience, and its components read
\begin{multline}
	\left(M^1\right)^{\{B_i,k_i\}}_{\{b_i,i_i\}}\equiv
	\sum_{\{j_i\}}
	\sqrt{\frac{d_{b_1}d_{b_2}d_{b_4}d_{b_5}d_{B_3}d_{B_6}}{d_{a_2}d_{a_5}d_{b_3}d_{b_6}d_A^2}}\\
	\left(\!\F{1}^{a_1b_1A}_{B_6}\!\right)^{B_1, j_1k_1}_{b_6,i_1j_6}
	\left(\!\Fi{1}^{b_1b_2B_2}_{B_1}\!\right)^{A,j_1j_2}_{a_2,i_2k_2}
	\left(\!\F{1}^{b_2a_3B_3}_{A}\!\right)^{B_2,k_3j_2}_{b_3,i_3j_3}\\
	\left(\!\Fi{1}^{b_4a_4B_3}_{A}\!\right)^{B_4,k_4j_4}_{b_3,i_4j_3}
	\left(\!\F{1}^{b_5b_4B_4}_{B_5}\!\right)^{A,j_4j_5}_{a_5,i_5k_5}
	\left(\!\Fi{1}^{a_6b_5A}_{B_6}\!\right)^{B_5,j_5k_6}_{b_6,i_6j_6}.
	\label{eq:Sublattice1Gs}
\end{multline}
For a plaquette of the second kind, we get
\begin{equation*}
	\input{sublattice_2_a.tikz}
	\hspace{-1.2em}=\hspace{-1.2em}\sum_{\{E_i,\alpha_i,h_i\}}\hspace{-1.2em}\!M^2\hspace{-1.2em}
	\input{sublattice_2_b.tikz}\hspace{-0.7em},
\end{equation*}
where the matrix $M^2$ depends on $\{C_i,D\}$ and its components read
\begin{multline}
	\left(M^2\right)^{\{E_i,\alpha_i,h_i\}}_{\{a_i,B_i,k_i\}}\equiv
	\sum_{\{l_i\}}
	\sqrt{\frac{d_{a_1}d_{a_2}d_{a_3}d_{B_2}d_{E_1}d_{\alpha_1}d_{\alpha_2}d_{\alpha_3}}{d_{C_1}d_{C_3}d_{C_4}d_{C_6}d_D^4}}\\
	\left(\!\Fi{2}^{a_1B_1\alpha_1}_{E_1}\!\right)^{D,l_2l_1}_{C_1,k_1h_1}
	\left(\!\Fi{2}^{a_2C_2\alpha_1}_{D}\!\right)^{E_2,h_2l_3}_{B_1,k_2l_2}
	\left(\!\Fi{2}^{a_2E_2\alpha_2}_{B_2}\!\right)^{C_3,h_3k_3}_{D,l_3l_4}\\
	\left(\!\F{2}^{a_3E_3\alpha_2}_{B_2}\!\right)^{C_4,h_4k_4}_{D,l_5l_4}
	\left(\!\F{2}^{a_3C_5\alpha_3}_{D}\!\right)^{E_3,h_5l_5}_{B_3,k_5l_6}
	\left(\!\F{2}^{a_1B_3\alpha_3}_{E_1}\!\right)^{D,l_6l_1}_{C_6,k_6h_6}.
	\label{eq:Sublattice2Gs}
\end{multline}
Finally, for the third kind of plaquette we get
\begin{equation*}
	\input{sublattice_3_a.tikz}
	\hspace{-1.2em}=\hspace{-0.7em}\sum_{\{\beta_i,o_i\}} \hspace{-0.7em} M^3\hspace{-1.2em}
	\input{sublattice_3_b.tikz}\hspace{-0.7em},
\end{equation*}
where the matrix $M^3$ depends on $\{\alpha_i,G\}$ and its components read
\begin{multline}
	\left(M^3\right)^{\{\beta_i,o_i\}}_{\{E_i,h_i\}}\equiv
	\sum_{\left\{n_i\right\}}
	\sqrt{\frac{d_{\beta_1}d_{\beta_2}d_{\beta_4}d_{\beta_5}d_{E_3}d_{E_6}}{d_{\alpha_2}d_{\alpha_5}d_{\beta_3}d_{\beta_6}d_G^2}}\\
	\left(\!\F{3}^{E_6\alpha_1\beta_1}_G\!\right)^{\beta_6,o_1n_6}_{E_1,h_1n_1}
	\left(\!\Fi{3}^{E_1\beta_1\beta_2}_{E_2}\!\right)^{\alpha_2,o_2h_2}_{G,n_1n_2}
	\left(\!\F{3}^{G\beta_2\alpha_3}_{E_3}\!\right)^{\beta_3,o_3n_3}_{E_2,n_2h_3}\\
	\left(\!\Fi{3}^{G\beta_4\alpha_4}_{E_3}\!\right)^{\beta_3,o_4n_3}_{E_4,n_4h_4}
	\left(\!\F{3}^{E_5\beta_5\beta_4}_{E_4}\!\right)^{\alpha_5,o_5h_5}_{G,n_5n_4}
	\left(\!\Fi{3}^{E_6\alpha_6\beta_5}_G\!\right)^{\beta_6,o_6n_6}_{E_5,h_6n_5}.
	\label{eq:Sublattice3Gs}
\end{multline}

The explicit expression for the operator $\left(p_{00}^{0,\mc{M}}\right)^{\otimes N}$ is then obtained by taking the appropriate linear combinations of the matrices $M^i$, and acting on their respective sublattices in the order described above. While the local Hilbert space of the two string-net models looks different, they have the same number of ground states, and therefore $\left(p_{00}^{0,\mc{M}}\right)^{\otimes N}$ is represented as a square matrix. To show that it is also unitary, we note that 
\begin{equation}
	\left(p_{00}^{0,\mc{M}}\right)^\dag p_{00}^{0,\mc{M}} = p_{00}^{0,\mc{D}_1},
\end{equation}
implying that the Hermitian conjugate of the matrix representation of $p_{00}^{0,\mc{M}}$ should correspond to the matrix representation of $\left(p_{00}^{0,\mc{M}}\right)^\dag$. By virtue of eq. \eqref{eq:ConjugateTube},
\begin{equation}
	\left(p_{00}^{0,\mc{M}}\right)^\dag = \frac{1}{D} \sum_{A} d_{A} \mc{B}_p^{A,\overline{\mc{M}}},
\end{equation}
which implies that the matrix representation of $\left(p_{00}^{0,\mc{M}}\right)^\dag$ is obtained as a linear combination of matrices $\overline{M}^i$, obtained in similar fashion as the matrices $M^i$. Using the fact that the different $F$-symbols themselves are unitary matrices since we are dealing with unitary bimodule categories, one can indeed show that $\left(M^i\right)^\dag = \overline{M}^i$ through a cumbersome but straightforward computation. Taking everything together, this implies that $\left(p_{00}^{0,\mc{M}}\right)^{\otimes N}$ is represented as a unitary matrix on the ground state.

\subsection{Excited states}
When applied to a state with excitations, the mapping provided by our circuit is less obvious since, as stressed before, the dimension of the Hilbert spaces does not match between the string-net model $\mc{D}_1$ and $\mc{D}_2$. We compensated for this by introducing an ancilliary qudit in each plaquette, that we initialize to be in the $\ket{0}$ state. Looking at the action of $V_{\mc{D}_1}^{\mc{D}_2}$ on the space spanned by $\ket{\psi^{\mc{D}_1},0}$, we find 
\begin{equation}
	V_{\mc{D}_1}^{\mc{D}_2} \ket{\psi^{\mc{D}_1},0} = \left(\sum_{a,j} p_{0j}^{a,\mc{M}} \otimes\ketbra{j}{0}\right)^{\otimes N}\ket{\psi^{\mc{D}_1},0}.
\end{equation}
Acting on any plaquette of the resulting state with a simple idempotent $p_{ii}^{a,\mc{D}_2}$ only gives a nonzero result if $i = 0$, meaning that for every central idempotent $P_a$ we only get the excitation corresponding to $p_{00}^a$. Additionally, this state is no longer part of the original Hilbert space spanned by $\ket{\psi^{\mc{D}_2}}$, as the ancilla $\ket{j}$ is now entangled with the other degrees of freedom of the string-net model $\mc{D}_2$.\\

There are a number of ways to disentangle these ancillas, but due to the mismatch in Hilbert space it is impossible to do so in a unique or unitary way. One option is to act with the operator $V_{\mc{D}_2}^{\mc{D}_2}$, which gives
\begin{align}
	V_{\mc{D}_2}^{\mc{D}_2} V_{\mc{D}_1}^{\mc{D}_2} \ket{\psi^{\mc{D}_1},0} &= \bigg(\sum_a \sum_{i=0}^{m_a-1} p_{ii}^{a,\mc{M}}\otimes\ketbra{0}\bigg)^{\otimes N}\ket{\psi^{\mc{D}_1},0}\nonumber\\
	&= \ket{\psi^{\mc{D}_2},0},
\end{align}
where $m_a = \min(n_a^{\mc{D}_1},n_a^{\mc{D}_2})$. This state is again part of the original Hilbert space, and the ancilla can be discarded. If $m_a = n_a^{\mc{D}_1}$, not all excitations of type $a \in Z(\mc{D}_2)$ in the string-net model $\mc{D}_2$ are in the image of this map, while if $m_a = n_a^{\mc{D}_2}$ some excitations of the string-net model $\mc{D}_1$ are projected out, all depending on the choice of enumeration of the simple idempotents. A different approach might be to make sure the energy of the state is conserved when mapping back to the original Hilbert space; we leave the exploration of the different possibilities to future work.\\

An explicit expression generalizing the matrices $M^i$ to the case of excitations is given in Appendix~\ref{App:ExtendedCircuit}. Unitarity of the full circuit can again be shown from unitarity of the $F$-symbols and the fact that the Hermitian conjugate of the bimodule tubes defined in eq. \eqref{eq:ConjugateTube}  coincides with the Hermitian conjugate of the matrix representation of the corresponding operator, by definition.

\section{Examples}
To illustrate the simple idempotents and simple bimodules that build up our circuit, we write them down for the case of a constant depth circuit that maps the toric code to itself. While this example is almost trivial, it provides enough insight to understand the more general case. We also present the required categorical data for the generic case of a quantum double.
\subsection{Toric code to toric code}
The toric code \cite{kitaev2003fault} is built from the fusion category $\mc{D}_1 = \text{Vec}_{\mathbb{Z}_2}$, the category of $\mathbb{Z}_2$ graded vector spaces with two simple objects $\{0,1\}$ that fuse according to $\mathbb Z_2$. It only hase one non-trivial module category $\mc{M}=\text{Vec}$, the category of finite dimensional vector spaces with a single simple object $\mathbb C$, from which one can construct an invertible bimodule category to $\mc{D}_2 = \text{Rep}(\mathbb{Z}_2) \simeq \text{Vec}_{\mathbb{Z}_2}$. We can use this bimodule category to construct a constant depth quantum circuit that maps the toric code to itself, but is not completely trivial. The relevant tubes of this bimodule category can be written as
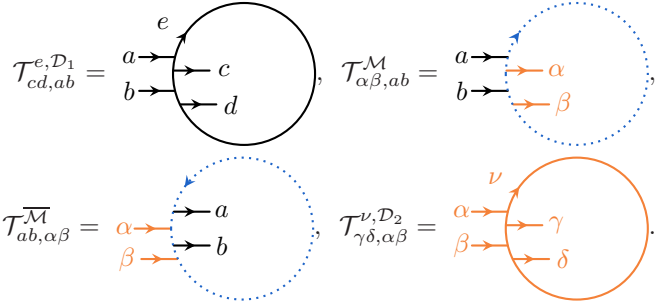
\begin{equation*}
	\begin{split}
		\mc{T}_{cd,ab}^{e,\mc{D}_1} = \toricTube{a}{b}{c}{d}{e}{D1}{D1}{D1}{1}, \enspace \mc{T}_{\alpha\beta,ab}^{\mc{M}} = \toricTube{a}{b}{\alpha}{\beta}{}{D1}{D2}{M}{1},\\
		\mc{T}_{ab,\alpha\beta}^{\overline{\mc{M}}} = \toricTube{\alpha}{\beta}{a}{b}{}{D2}{D1}{M}{-1}, \enspace \mc{T}_{\gamma\delta,\alpha\beta}^{\nu,\mc{D}_2} = \toricTube{\alpha}{\beta}{\gamma}{\delta}{\nu}{D2}{D2}{D2}{1}.
	\end{split}
\end{equation*}
The central idempotents with their corresponding simple idempotents and nilpotents of the tube algebra of $\mc{D}_1$ are labeled by the four excitations in the toric code. These are the vaccuum $\mathbb 1$:
\begin{equation*}
	\begin{split}
		p_{00}^{\mathbb{1},\mc{D}_1} &= \frac{1}{2}\left(\mc{T}_{00,00}^{0,\mc{D}_1} + \mc{T}_{00,00}^{1,\mc{D}_1}\right), \enspace p_{01}^{\mathbb{1},\mc{D}_1} = \frac{1}{2}\left(\mc{T}_{00,11}^{0,\mc{D}_1} + \mc{T}_{00,11}^{1,\mc{D}_1}\right),\\
		p_{10}^{\mathbb{1},\mc{D}_1} &= \frac{1}{2}\left(\mc{T}_{11,00}^{0,\mc{D}_1} + \mc{T}_{11,00}^{1,\mc{D}_1}\right), \enspace p_{11}^{\mathbb{1},\mc{D}_1} = \frac{1}{2}\left(\mc{T}_{11,11}^{0,\mc{D}_1} + \mc{T}_{11,11}^{1,\mc{D}_1}\right),
	\end{split}
\end{equation*}
the electric excitation $e$:
\begin{equation*}
	\begin{split}
		p_{00}^{e,\mc{D}_1} &= \frac{1}{2}\left(\mc{T}_{01,01}^{0,\mc{D}_1} + \mc{T}_{01,01}^{1,\mc{D}_1}\right), \enspace p_{01}^{e,\mc{D}_1} = \frac{1}{2}\left(\mc{T}_{01,10}^{0,\mc{D}_1} + \mc{T}_{01,10}^{1,\mc{D}_1}\right),\\
		p_{10}^{e,\mc{D}_1} &= \frac{1}{2}\left(\mc{T}_{10,01}^{0,\mc{D}_1} + \mc{T}_{10,01}^{1,\mc{D}_1}\right), \enspace p_{11}^{e,\mc{D}_1} = \frac{1}{2}\left(\mc{T}_{10,10}^{0,\mc{D}_1} + \mc{T}_{10,10}^{1,\mc{D}_1}\right),
	\end{split}
\end{equation*}
the magnetic excitation $m$:
\begin{equation*}
	\begin{split}
		p_{00}^{m,\mc{D}_1} &= \frac{1}{2}\left(\mc{T}_{00,00}^{0,\mc{D}_1} - \mc{T}_{00,00}^{1,\mc{D}_1}\right), \enspace p_{01}^{m,\mc{D}_1} = \frac{1}{2}\left(\mc{T}_{00,11}^{0,\mc{D}_1} - \mc{T}_{00,11}^{1,\mc{D}_1}\right)\\
		p_{10}^{m,\mc{D}_1} &= \frac{1}{2}\left(\mc{T}_{11,00}^{0,\mc{D}_1} - \mc{T}_{11,00}^{1,\mc{D}_1}\right), \enspace p_{11}^{m,\mc{D}_1} = \frac{1}{2}\left(\mc{T}_{11,11}^{0,\mc{D}_1} - \mc{T}_{11,11}^{1,\mc{D}_1}\right),
	\end{split}
\end{equation*}
and the fermion $f$:
\begin{equation*}
	\begin{split}
		p_{00}^{f,\mc{D}_1} &= \frac{1}{2}\left(\mc{T}_{01,01}^{0,\mc{D}_1} - \mc{T}_{01,01}^{1,\mc{D}_1}\right), \enspace p_{01}^{f,\mc{D}_1} = \frac{1}{2}\left(\mc{T}_{01,10}^{0,\mc{D}_1} - \mc{T}_{01,10}^{1,\mc{D}_1}\right)\\
		p_{10}^{f,\mc{D}_1} &= \frac{1}{2}\left(\mc{T}_{10,01}^{0,\mc{D}_1} - \mc{T}_{10,01}^{1,\mc{D}_1}\right), \enspace p_{11}^{f,\mc{D}_1} = \frac{1}{2}\left(\mc{T}_{10,10}^{0,\mc{D}_1} - \mc{T}_{10,10}^{1,\mc{D}_1}\right).
	\end{split}
\end{equation*}
These central idempotents are all built from two simple idempotents, reflecting the fact that these tube algebras describe the excitations of two vertices and one plaquette. For instance, the simple idempotent $p_{00}^{\mathbb{1},\mc{D}_1}$ corresponds to the true vacuum, where neither the vertices nor the plaquette is excited. The other simple idempotent $p_{11}^{\mathbb{1},\mc{D}_1}$ represents the case where both vertices are excited; together, they are in the vacuum, and therefore they also contribute to the vacuum central idempotent. The other central idempotents follow a similar pattern. The simple idempotents and nilpotents of the tube algebra of $\mc{D}_2$ are related to those of the tube algebra of $\mc{D}_1$ as follows:
\begin{equation*}
	\begin{split}
		p_{ij}^{\mathbb{1},\mc{D}_2} = p_{ij}^{\mathbb{1},\mc{D}_1}, \enspace p_{ij}^{e,\mc{D}_2} = p_{ij}^{m,\mc{D}_1}\\
		p_{ij}^{m,\mc{D}_2} = p_{ij}^{e,\mc{D}_1}, \enspace p_{ij}^{f,\mc{D}_2} = p_{ij}^{f,\mc{D}_1}.
	\end{split}
\end{equation*}
We note that the electric and magnetic excitations are swapped, which is a manifestation of the $e-m$ duality in the toric code. From these idempotents and nilpotents, one can compute the simple bimodules of $\mc{M}$. We find
\begin{equation*}
	\begin{split}
		p_{00}^{\mathbb{1},\mc{M}} &= \frac{1}{\sqrt{2}}\mc{T}_{00,00}^{\mc{M}}, \enspace p_{01}^{\mathbb{1},\mc{M}} = \frac{1}{\sqrt{2}}\mc{T}_{00,11}^{\mc{M}}\\
		p_{10}^{\mathbb{1},\mc{M}} &= \frac{1}{\sqrt{2}}\mc{T}_{11,00}^{\mc{M}}, \enspace p_{11}^{\mathbb{1},\mc{M}} = \frac{-1}{\sqrt{2}}\mc{T}_{11,11}^{\mc{M}}
	\end{split}
\end{equation*}
for the vacuum,
\begin{equation*}
	\begin{split}
		p_{00}^{e,\mc{M}} &= \frac{1}{\sqrt{2}}\mc{T}_{00,01}^{\mc{M}}, \enspace p_{01}^{e,\mc{M}} = \frac{1}{\sqrt{2}}\mc{T}_{00,10}^{\mc{M}}\\
		p_{10}^{e,\mc{M}} &= \frac{-1}{\sqrt{2}}\mc{T}_{11,01}^{\mc{M}}, \enspace p_{11}^{e,\mc{M}} = \frac{1}{\sqrt{2}}\mc{T}_{11,10}^{\mc{M}}
	\end{split}
\end{equation*}
for the electric excitation,
\begin{equation*}
	\begin{split}
		p_{00}^{m,\mc{M}} &= \frac{1}{\sqrt{2}}\mc{T}_{01,00}^{\mc{M}}, \enspace p_{01}^{m,\mc{M}} = \frac{1}{\sqrt{2}}\mc{T}_{01,11}^{\mc{M}}\\
		p_{10}^{m,\mc{M}} &= \frac{1}{\sqrt{2}}\mc{T}_{10,00}^{\mc{M}}, \enspace p_{11}^{m,\mc{M}} = \frac{-1}{\sqrt{2}}\mc{T}_{10,11}^{\mc{M}}
	\end{split}
\end{equation*}
for the magnetic excitation and
\begin{equation*}
	\begin{split}
		p_{00}^{f,\mc{M}} &= \frac{1}{\sqrt{2}}\mc{T}_{01,01}^{\mc{M}}, \enspace p_{01}^{f,\mc{M}} = \frac{1}{\sqrt{2}}\mc{T}_{01,10}^{\mc{M}}\\
		p_{10}^{f,\mc{M}} &= \frac{-1}{\sqrt{2}}\mc{T}_{10,01}^{\mc{M}}, \enspace p_{11}^{f,\mc{M}} = \frac{1}{\sqrt{2}}\mc{T}_{10,10}^{\mc{M}}
	\end{split}
\end{equation*}
for the fermion. The simple bimodules of $\overline{\mc{M}}$ are obtained by Hermitian conjugation of the simple bimodules of $\mc{M}$:
\begin{equation*}
	\begin{split}
		p_{ji}^{\mathbb{1},\overline{\mc{M}}} &= \left(p_{ij}^{\mathbb{1},\mc{M}}\right)^\dag, \enspace p_{ji}^{e,\overline{\mc{M}}} = \left(p_{ij}^{e,\mc{M}}\right)^\dag,\\
		p_{ji}^{m,\overline{\mc{M}}} &= \left(p_{ij}^{m,\mc{M}}\right)^\dag, \enspace p_{ji}^{f,\overline{\mc{M}}} = \left(p_{ij}^{f,\mc{M}}\right)^\dag.
	\end{split}
\end{equation*}

\subsection{Ground states of Kitaev's quantum doubles to string-nets}

The previous example of the toric code can now be generalized to a circuit mapping between quantum doubles and string-net ground states. In those examples the group $\mb{Z}_2$ is replaced by an arbitrary finite group $G$. This then amounts to choosing the UFC $\mc{D}_1$ to be $\mc{D}_1=\text{Vec}_\text{G}$, the category of all finite dimensional $G$-graded vector spaces which is the input category of Kitaev's quantum doubles~\cite{kitaev2003fault}. The simple objects of $\text{Vec}_\text{G}$ are in one-to-one correspondence with the group elements of $G$ such that the fusion of elements in $\text{Vec}_G$ is then equivalent to the group multiplication in $G$. The ground states of these quantum doubles can then be mapped to string-nets with input category $\mc{D}_2=\text{Rep}(G)$, the category of all representations of $G$ in which the simple objects are the irreducible representations. The invertible bimodule category connecting $\mc{D}_1$ and $\mc{D}_2$ is $\mc{M}=\text{Vec}$, as in the case of the toric code. A crucial difference with the previous example is that for a generic group, there is no monoidal equivalence between $\text{Vec}_\text{G}$ and $\text{Rep}(G)$. The data of this bimodule category is as follows~\cite{lootens2021matrix}:

\begin{itemize}
	\item $\left(\F{0}^{g_1,g_2,g_3}_{g_{123}}\right)^{g_{23},11}_{g_{12},11}=\omega(g_1,g_2,g_3)$ where $\omega$ is a 3-cocycle belonging to the trivial class. One can choose a gauge of the $\F{0}$ symbol in which this cocycle is identically one, $\omega(g_1,g_2,g_3)\equiv 1, \forall g_1,g_2,g_3\in G$.
	\item Writing $\left(\F{1}^{g_1g_2\mb{C}}_\mb{C}\right)^{\mb{C},11}_{g_{12},11}=\psi(g_1,g_2)$, it can be shown that by virtue of the pentagon equation between $\F{0}$ and $\F{1}$, $\psi$ is a 2-cocycle classified by $H^2(G,U(1))$.
	\item $\left(\F{2}^{g\mb{C}\alpha}_\mb{C}\right)^{\mb{C},j1}_{\mb{C},1i} = D^\alpha(g)^j_i$ where $D^\alpha(g)$ is the matrix representation of the group element $g\in\text{G}$ in the irreducible representation $\alpha$ and where $i,j$ are $d_\alpha$-dimensional indices in that representation.
	\item $\left(\F{3}^{\mb{C}\alpha_1\alpha_2}_\mb{C}\right)^{\alpha_3,ki_3}_{\mb{C},i_1i_2}=C^{\alpha_1\alpha_2\alpha_3}_{i_1i_2i_3,k}$ are the Clebsch-Gordan coefficients of the group. In this notation $k$ denotes the possible number of ways in which $\alpha_1$ and $\alpha_2$ can fuse to $\alpha_3$.
	\item The $\F{4}$-symbols can then be recognized as the Racah W-coefficients which are equal to the $6j$ symbols of $G$. These $6j$ symbols serve as the input of the $\text{Rep}(G)$ string-net.
\end{itemize}

Substituting these F-symbols in the expressions of the quantum circuit then results in a circuit that allows one to map the ground states of all of Kitaev's quantum double models to corresponding string-nets. A similar circuit was found in \cite{buerschaper2009mapping, kadar2010microscopic}. In their map the Rep($G$) degrees-of-freedom where introduced by means of a Fourier transform on $G$. In our circuit these irreps appear through the $\F{2}$-symbol of the bimodele category. We also recover the $6j$ symbols that appeared in their construction in the form of the $\F{4}$-symbols. We note that the quantum doubles as defined in \cite{buerschaper2009explicit,kadar2010microscopic} are defined on a triangular lattice and are shown to be equivalent to the string-net model on the hexagonal lattice; our construction does not change the lattice, and a direct comparison is therefore somewhat cumbersome.

\section{Conclusion and outlook}
In this work, we have constructed a constant depth quantum circuit that is able to map between Morita equivalent ground states. Crucially, the circuit is unitary on the full Hilbert space, meaning that it can be implemented in a truly local way without performing a measurement as would be required to construct the string-net state from a product state. We have shown that the mapping is exact for the ground states, which implies that they are in the same phase as they can be adiabatically connected without closing a gap.\\

One immediate application of our circuit exploits the recent insight that string-net models can be used to construct partition functions of classical statistical mechanics models \cite{aasen2016topological,aasen2020topological}. From the tensor network point of view, this is understood as the strange correlator \cite{vanhove2018mapping}, which boils down to taking the overlap of the string-net ground state PEPS with some unentangled product state to obtain a tensor network representation of a classical partition function. The existence of a quantum circuit between Morita equivalent string-nets allows one to construct different lattice models with the same partition functions. We plan to explore this further in future work.\\

A natural generalization would be to no longer restrict to closed manifolds but also include the case of boundaries to the vacuum or domain walls between different string-net models. In general this requires the use of a 4-object bicategory containing 4 fusion categories and 6 invertible bimodule categories \cite{lootens2021matrix}, the recoupling theory of which has not been written down explicitly in the literature.

\section*{Acknowledgments}
We are grateful to Nick Bultinck for inspiring discussions, and Jacob Bridgeman for insightful comments on invertible bimodule categories and tube algebras. This work has received funding from the European Research Council (ERC) under the European Union’s Horizon 2020 research and innovation  programme  (grant agreements No 647905 (QUTE) and No 863476 (SEQUAM)), and the Research Foundation Flanders via grant nr.\ G087918N and G0E1820N. LL is supported by a PhD fellowship from the Research Foundation Flanders (FWO). BVDC is supported by a PhD fellowship from Bijzonder Onderzoeksfonds (BOF).

\onecolumngrid
\appendix

\section{The quantum circuit for the extended string-net}
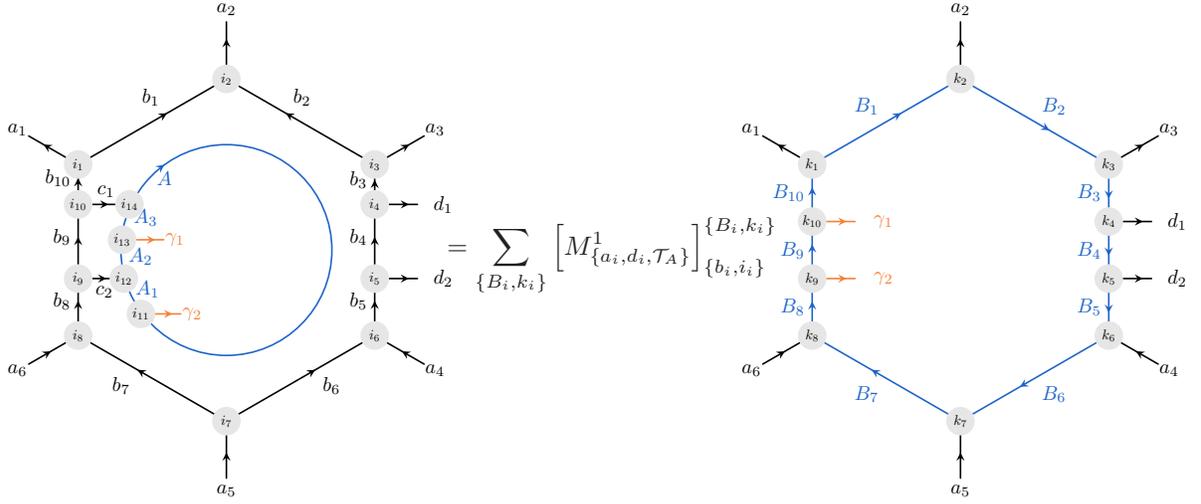
\label{App:ExtendedCircuit}

\begin{equation*}
	\begin{gathered}
		\input{sublattice_1_a_extended.tikz}
		\hspace{-10pt}
		=\sum_{\left\{B_i,k_i\right\}}
		\left[M^{1}_{\left\{a_i,d_i,\mc{T}_A\right\}}\right]^{\{B_i,k_i\}}_{\{b_i,i_i\}}
		\hspace{-20pt}
		\input{sublattice_1_b_extended.tikz}
		\\
		\left[M^{1}_{\{a_i,d_i,\mc{T}_A\}}\right]^{\{B_i,k_i\}}_{\{b_i,i_i\}}\equiv
		\sum_{\{j_i\}}
		\left(\frac{d_{b_1}d_{b_2}d_{b_6}d_{b_7}d_{c_1}d_{c_2}d_{A_1}d_{A_3}d_{B_5}d_{B_8}}{d_A^3d_{a_2}d_{a_5}d_{b_5}d_{b_8}d_{A_2}}\right)^{1/2}
		\left(\F{1}^{b_{10}c_1A_3}_{B_{10}}\right)^{A_1,i_{14}j_{12}}_{b_9,i_{10}j_{10}}
		\left(\Fi{2}^{b_9A_3\gamma_1}_{B_9}\right)^{A_2,i_{13}j_{10}}_{B_{10},j_{11}k_{10}}
		\\\times
		\left(\F{1}^{b_9c_2A_1}_{B_9}\right)^{A_2,i_{12}j_{10}}_{b_8,i_9j_9}
		\left(\Fi{2}^{b_8A_1\gamma_2}_{B_8}\right)^{A_3,i_{11}j_8}_{B_9,j_9k_9}
		\left(\F{1}^{a_1b_1A}_{B_{10}}\right)^{B_1, j_1k_1}_{b_{10},i_1j_{12}}
		\left(\Fi{1}^{b_1b_2B_2}_{B_1}\right)^{A,j_1j_2}_{a_2,i_2k_2}
		\left(\F{1}^{b_2a_3B_3}_{A}\right)^{B_2,k_3j_2}_{b_3,i_3j_3}
		\\\times
		\left(\F{1}^{b_4d_2B_5}_A\right)^{B_4,k_5j_4}_{b_5,i_5j_5}
		\left(\F{1}^{b_3d_1B_4}_A\right)^{B_3,k_4j_3}_{b_4,i_4j_4}
		\left(\Fi{1}^{b_6a_4B_5}_{A}\right)^{B_6,k_6j_6}_{b_5,i_6j_5}
		\left(\F{1}^{b_7b_6B_6}_{B_7}\right)^{A,j_6j_7}_{a_5,i_7k_7}
		\left(\Fi{1}^{a_6b_7A}_{B_8}\right)^{B_7,j_7k_8}_{b_8,i_8j_8}
	\end{gathered}
\end{equation*}

\begin{equation*}
	\begin{gathered}
		\input{sublattice_2_a_extended.tikz}
		\hspace{-10pt}
		=\sum_{\left\{E_i,\alpha_i,m_i\right\}}
		\left[M^{2}_{\left\{C_i,\gamma_i,\mc{T}_D\right\}}\right]^{\{E_i,\alpha_i,m_i\}}_{\{B_i,a_i,k_i\}}
		\hspace{-20pt}
		\input{sublattice_2_b_extended.tikz}
		\\
		\left[M^{2}_{\left\{C_i,\gamma_i,\mc{T}_D\right\}}\right]^{\{E_i,\alpha_i,m_i\}}_{\{B_i,a_i,k_i\}}
		\hspace{-3pt}
		\equiv
		\hspace{-3pt}
		\sum_{\{l_i\}}
		\hspace{-2pt}
		\left(\frac{d_{a_1}d_{a_2}d_{a_5}d_{c_1}d_{c_2}d_{D_1}d_{D_3}d_{B_4}d_{E_3}d_{E_4}d_{\alpha_1}d_{\alpha_2}d_{\alpha_5}}{d_D^5d_{C_1}d_{C_3}d_{C_4}d_{C_6}d_{D_2}d_{E_4}}\right)^{1/2}
		\hspace{-6pt}
		\left(\F{1}^{a_5c_1D_3}_{E_5}\right)^{D,k_{14}l_{12}}_{a_4,k_{10}l_{11}}
		\hspace{-4pt}
		\left(\Fi{2}^{a_4D_3\beta_1}_{E_4}\right)^{D_2,k_{13}l_{10}}_{E_5,l_{11}m_{10}}
		\\\times
		\left(\F{1}^{a_4c_2D_1}_{E_4}\right)^{D_2,k_{12}l_{10}}_{a_3,k_9l_9}
		\left(\Fi{2}^{a_3D_1\beta_2}_{E_3}\right)^{D,k_{11}l_8}_{E_4,l_9m_9}
		\left(\Fi{2}^{a_5B_1\alpha_1}_{E_5}\right)^{D,l_1l_{12}}_{C_1,k_1m_1}
		\left(\Fi{2}^{a_1C_2\alpha_1}_{D}\right)^{E_1,m_2l_2}_{B_1,k_2l_1}
		\left(\Fi{2}^{a_1E_1\alpha_2}_{B_2}\right)^{C_3,m_3k_3}_{D,l_2l_3}
		\\\times
		\left(\F{3}^{D\alpha_2\gamma_1}_{B_3}\right)^{\alpha_3,m_4l_4}_{B_2,l_3k_4}
		\left(\F{3}^{D\alpha_3\gamma_2}_{B_4}\right)^{\alpha_4,m_5l_5}_{B_3,l_4k_5}
		\left(\F{2}^{a_2E_2\alpha_4}_{B_4}\right)^{C_4,m_6k_6}_{D,l_6l_5}
		\left(\F{2}^{a_2C_5\alpha_5}_{D}\right)^{E_2,m_7l_6}_{B_5,k_7l_7}
		\left(\F{2}^{a_3B_5\alpha_5}_{E_3}\right)^{D,l_7l_8}_{C_6,k_8m_8}
	\end{gathered}
\end{equation*}

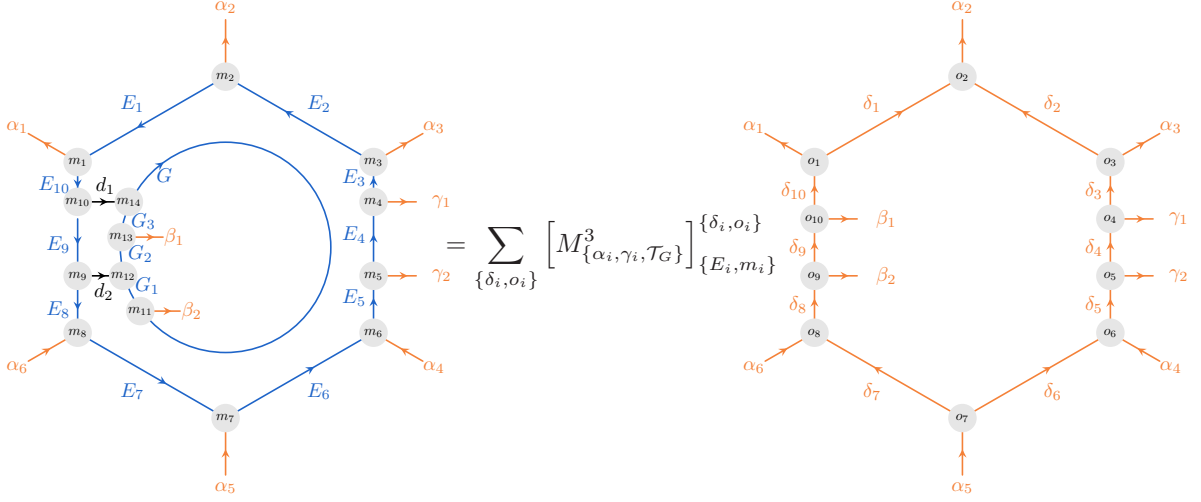
\begin{equation*}
	\begin{gathered}
		\input{sublattice_3_a_extended.tikz}
		\hspace{-10pt}
		=\sum_{\left\{\delta_i,o_i\right\}}
		\left[M^{3}_{\left\{\alpha_i,\gamma_i,\mc{T}_G\right\}}\right]^{\{\delta_i,o_i\}}_{\{E_i,m_i\}}
		\hspace{-20pt}
		\input{sublattice_3_b_extended.tikz}
		\\
		\left[M^{3}_{\left\{\alpha_i,\gamma_i,\mc{T}_G\right\}}\right]^{\{\delta_i,o_i\}}_{\{E_i,m_i\}}\equiv
		\sum_{\{n_i\}}
		\left(\frac{d_{d_1}d_{d_2}d_{E_1}d_{E_2}d_{E_3}d_{E_5}d_{E_6}d_{E_7}d_{E_8}^2d_{E_{10}}d_{G_2}d_{\delta_1}d_{\delta_2}d_{\delta_6}d_{\delta_7}}
		{d_G^2d_{G_1}d_{G_3}d_{\alpha_2}d_{\alpha_5}d_{\delta_8}}\right)^{1/2}
		\left(\F{2}^{d_1E_9\delta_{10}}_{G}\right)^{G_3,n_9m_{14}}_{E_{10},m_{10}n_{10}}
		\\\times
		\left(\F{3}^{E_9\delta_{10}\beta_1}_{G_2}\right)^{\delta_9,o_{10}n_8}_{G_3,n_9m_{13}}
		\left(\F{2}^{d_2E_8\delta_9}_{G_2}\right)^{G_1,n_{11}m_{12}}_{E_9,m_9n_{12}}
		\left(\F{3}^{E_8\delta_9\beta_{12}}_{G}\right)^{\delta_8,o_9n_{10}}_{G_1,n_{11}m_{11}}
		\left(\F{3}^{E_{10}\alpha_1\delta_1}_{G}\right)^{\delta_{10},o_1n_{12}}_{E_1,m_1n_1}
		\left(\Fi{3}^{E_1\delta_1\delta_2}_{E_2}\right)^{\alpha_2,o_2m_2}_{G,n_1n_2}
		\\\times
		\left(\F{3}^{G\delta_2\alpha_3}_{E_3}\right)^{\delta_3,o_3n_3}_{E_2,n_2m_3}
		\left(\F{3}^{G\delta_3\gamma_1}_{E_4}\right)^{\delta_4,o_4n_4}_{E_3,n_3m_4}
		\left(\F{3}^{G\delta_4\gamma_2}_{E_5}\right)^{\delta_5,o_5n_5}_{E_4,n_4m_5}
		\left(\Fi{3}^{G\delta_6\alpha_4}_{E_5}\right)^{\delta_5,o_6n_5}_{E_6,n_6m_6}
		\left(\F{3}^{E_7\delta_7\delta_6}_{E_6}\right)^{\alpha_5,o_7m_7}_{G,n_7n_6}
		\left(\Fi{3}^{E_8\alpha_6\delta_7}_{G}\right)^{\delta_8,o_8n_8}_{E_7,m_8n_7}
	\end{gathered}
\end{equation*}

\end{document}